\documentclass[reprint, superscriptaddress, secnumarabic, amssymb, nobibnotes, aps, prl]{revtex4-1}

\usepackage{graphicx}
\usepackage{epstopdf}
\usepackage[T1]{fontenc}
\usepackage{amsbsy}
\usepackage{gensymb}
\usepackage[T1]{fontenc}
\usepackage{amsmath}
\usepackage{amssymb}
\usepackage{bbm}
\usepackage{braket}
\usepackage{xcolor}
\allowdisplaybreaks
\usepackage{graphicx}
\usepackage[colorlinks=true]{hyperref}  
\hypersetup{
    bookmarks=true,         
    unicode=false,          
    pdftoolbar=true,        
    pdfmenubar=true,        
    pdffitwindow=false,     
    pdfstartview={FitH},    
    pdftitle={Investigation on superconductivity in Laves phase Re$_{2}$A (A = Hf, Zr)},    
    pdfauthor={M. Mandal, R. P. Singh},      
    pdfsubject={},   
    pdfcreator={},   
    pdfproducer={}, 
    pdfkeywords={} {} {}, 
    pdfnewwindow=true,      
    colorlinks=true,       
    linkcolor=blue, 
    citecolor=blue,        
    filecolor=magenta,      
    urlcolor=blue           
} 
\usepackage[normalem]{ulem}

\newcommand{\figref}[1]{Fig.~\ref{#1}}

\newcommand{\tableref}[1]{Table~\ref{#1}}

\renewcommand{\approx}{\simeq}

\begin{document}
\title{\textrm{Emergent superconductivity by Re doping in type -II Weyl semimetal NiTe$_{2}$}}
\author{M. Mandal}
\affiliation{Department of Physics, Indian Institute of Science Education and Research Bhopal, Bhopal, 462066, India}
\author{R. P. Singh}
\email[]{rpsingh@iiserb.ac.in}
\affiliation{Department of Physics, Indian Institute of Science Education and Research Bhopal, Bhopal, 462066, India}
\date{\today}
\begin{abstract}
\begin{flushleft}

\end{flushleft}
Recently topological superconductors emerge as a platform to realise Majorana Fermions which has potential application in topological quantum computation. Superconductivity in these materials can be realised by chemical doping, high pressure or proximity effect. Herein we report the emergent superconductivity in single-crystals of Re doped type-II Weyl semimetal NiTe$_{2}$. The magnetic and transport measurements highlight that Re substitution in Ni-site induces superconductivity at a maximum temperature of 2.36 K. Hall effect and specific heat measurements indicate that Re substitution is doping hole and facilitates the emergence of superconductivity by phonon softening and enhancing the electron-phonon coupling.

\end{abstract}
\maketitle

\section{Introduction}

Following the discovery of topological insulators, i.e., a new state of quantum matter which is insulating in bulk, but exhibit unique surface states that are conducting and topologically protected-i.e.; the spin orbit coupling in these materials leads to the formation of surface states that cannot be destroyed by scattering or impurities, the search for quantum materials with distinct symmetry protected topological phenomena has become one of the main aspirations in condensed matter physics \cite{topology}. Thenceforth, multiple series of two-dimensional and three-dimensional topological semimetals (TSM) \cite{2DTSM2, 2DTSM3, 3DTSM, 3DTSM2}, including Dirac semimetals (DSM) and Weyl semimetals (WSM), are widely researched with theoretical models and experimental results. The discovery of the topological Weyl semimetals (WSMs) has sparked enormous research interests recently since it provides the direct realization of the Weyl fermions in condensed matter physics. Recently, the concept of broken Lorentz invariance is also introduced to WSM, resulting in a newly emerging system named type-II WSM, where the Dirac cones are greatly tilted in a momentum direction \citep{weyl1, weyl2}, leading to modulated effective mass and versatile device opportunities \cite{application1,application2,application3,application4}. In the case of coexisting superconductivity, topological superconductivity (TSC) \cite{topology_sc,topology_sc2,topology_sc3,topology_sc4, topology_sc5} may be expected with the potential future of Majorana fermion \cite{majorana1, majorana2} and topological quantum computation \cite{quantum_information1, quantum_information2} where broader momentum dispersion hosts more freedoms. The real materials of type-II WSM are discovered in 1T-phase PtSe$_{2}$ family, i.e. PtSe$_{2}$ \cite{PtSe2, PtSe2_2}, PtTe$_{2}$ \cite{PtTe2} and PdTe$_{2}$ \cite{PdTe2, PdTe2_2}, recently. All these materials are group-10 transition-metal dichalcogenides (TMDC) with similar structures and space group. In spite of large scale research in these materials, an issue was plagued that the Dirac points are far below the Fermi level, and there are several trivial bands crossing the Fermi surface. Since Dirac fermions close to the Fermi energy imply more prominent contributions from these relativistic carriers in its transport and thermodynamic properties, a type-II WSM system with Dirac points closer to Fermi level represent an improved platform to study the type-II Dirac physics. NiTe$_{2}$  is reported as  type -II Dirac semimetal with a pair Dirac nodes very close to the Fermi level \cite{NiTe2} compared with its homolog PtSe$_{2}$ \cite{PtSe2, PtSe2_2}, PtTe$_{2}$ \cite{PtTe2} and PdTe$_{2}$ \cite{PdTe2, PdTe2_2}.  Since the quantum oscillations in the same material reveal a nontrivial Berry’s phase associated with these Dirac fermions, it also promotes possible applications based on these topological carriers. Besides, as a special part of TMDs, most of them, MX$_{2}$ (M = Mo, W, Zr, Pt, Pd, Ni; X = Te ) can achieve superconductivity by doping, intercalation or pressure. Superconductivity in Weyl semimetal candidate MoTe$_{2}$ can be enhanced by partial replacing the Te with S \cite{S} and Se \cite{Se} and also by doping Re in Mo site \cite{Re}. Similarly, intercalation of potassium induces superconductivity in WTe$_{2}$ \cite{K}. Besides, the application of external pressure also induces the superconductivity in  MoTe$_{2}$ \cite{MoTe2_pressure} and  WTe$_{2}$ \cite{WTe2_pressure}. Very recently, NiTe$_{2}$ has been purposed as the possible topological superconductor with external pressure and T$_{c}$ reaches its maximum, 7.5 K, at the pressure of 52.8 GPa \cite{NiTe2_pressure}. A computational investigation based on the anisotropic Midgal-Eliashberg formalism reports monolayer NiTe$_{2}$ to be an intrinsic superconductor with a T$_{c}$ = 5.7 K, although the bulk crystal is not known to superconductor \cite{NiTe2_2}. In contrast to conventional superconductors, the generally negative influence of external pressure on T$_{c}$ is interpreted as the importance of the electron phonon interaction for triggering the phase transition \cite{book2}. So, it would be enthralling to inspect the possible topological superconductivity in this material by chemical substitution or intercalation.
Here, we present evidence that Re doped in NiTe$_{2}$ induces superconductivity. The dependence of superconducting transition temperature T$_{c}$ with a composition (x) and hole doping in the structure is described here. Furthermore, combining the ac and dc susceptibility, resistivity, specific heat, and hall measurements, we demonstrate the variation of superconducting properties with Re doping in NiTe$_{2}$ system at ambient pressure.
\section{Experimental}

Single crystals of Ni$_{1-x}$Re$_{x}$Te$_{2}$ (x = 0, 0.05, 0.1 and 0.2) were grown by the modified Bridgman method. In the first step, stoichiometric mixtures of Ni (99.95 $ \% $pure), (99.99 $ \% $) and Te (99.99$ \% $ pure) powders were ground together, pelletized and sealed in an evacuated quartz tube. The sealed ampoule was first heated at 700\degree C for 48 hours, followed by ice water quenching to avoid the formation of the impurity phase and the same heat treatment was repeated. The Re doped polycrystalline samples were used in the crystal growth process. Crystallization was carried out by slow cooling up to 700\degree C where the polycrystalline sample was kept at 900\degree C in a conical quartz tube. Finally, the sealed quartz tube was quenched in cold water to avoid the formation of the impurity phase.

The obtained samples were examined by PANalytical X$^{,}$pert Pro diffractometer equipped with Cu $K_{\alpha}$ radiation ($\lambda$ = 1.54056 $\text{\AA}$) at room temperature for the characterization of crystal structure and phase purity. Rietveld refinement was performed using Full Prof Suite Software. Compositions of the samples were checked by an energy-dispersive X-ray spectrometer (EDS) which confirms the existence of Re and Ni or Te in the samples in nominal quantity. The single-crystal orientation was checked by a standard Laue diffraction technique. To study the superconducting state, we measured DC and AC susceptibility measurement by Superconducting Quantum Interference Device (SQUID MPMS, Quantum Design). In magnetization measurement, the samples were cooled down to 1.8 K in zero fields and the data were recorded while warming it to 5 K in 10 mT field called ZFCW mode, whereas in the FCC mode, the sample was cooled down to 1.8 K in the presence of the same field with data taken simultaneously. Magnetization measurements were also done under different applied fields in a temperature range of 1.8 K to 5.0 K. Specific heat measurements were performed by the two tau time-relaxation method in zero fields in the Physical Property Measurement System (PPMS, Quantum Design, Inc.). The electrical resistivity measurements were performed by using conventional four-probe ac technique at frequency 157 Hz and excitation current 10 mA in zero fields from 1.85 K to 300 K to know the residual resistivity $\rho_{0}$ and residual resistivity ratio (RRR) of the same specimen. Hall measurements were carried out on the same system at frequency 157 Hz and excitation current 10 mA in the field range $\pm$ 8 T. 

\section{Results and discussion}

\begin{figure}[htbp!]
\includegraphics[width=1.0\columnwidth]{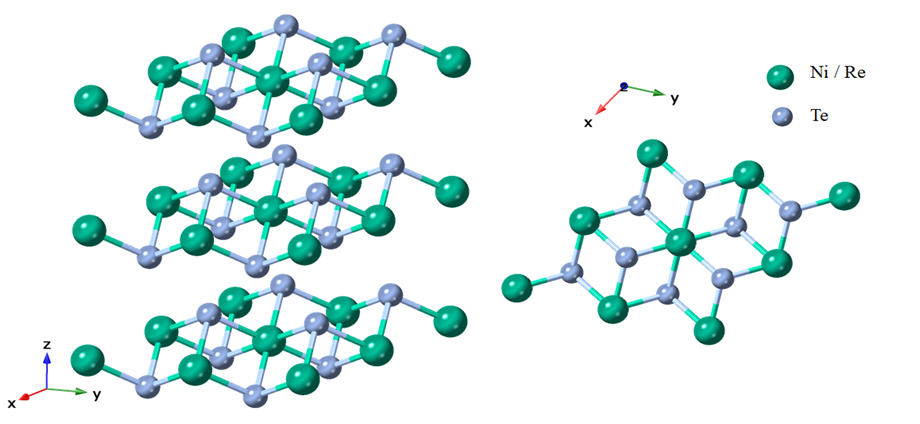}
\caption {\label{Fig1:crystal_structure} CdI$_{2}$ prototype crystal structure of NiTe$_{2}$ having the space group P-3m1 (164).}
\end{figure}

\begin{figure}[htbp!]
\includegraphics[width=1.0\columnwidth]{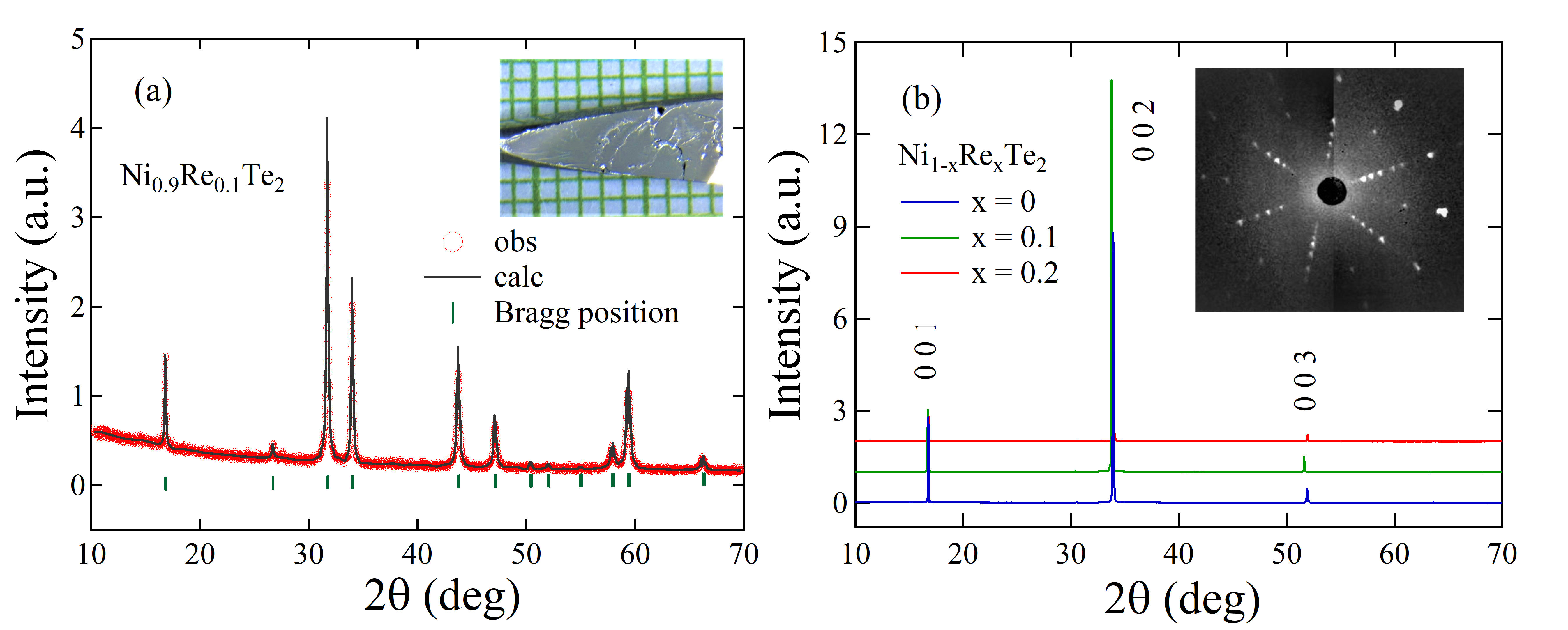}
\caption {\label{Fig2:xrd} (a) Rietveld refinement of polycrystaline sample Ni$_{0.9}$Re$_{0.1}$Te$_{2}$ whereas inset shows the crystal picture. (b) Single crystal growth in (00n) direction and Laue image is shown in the inset.}
\end{figure}

\begin{table}
  \caption{Average elemental concentration (average of 10 points) obtained
from the EDS measurements for Ni$_{1-x}$ Re$_{x}$Te$_{2}$ samples.}
  \label{tbl:EDS}
  \begin{tabular}{ll}
    \hline
    Nominal composition & From EDS  \\
    \hline
    NiTe$_{2}$   & NiTe$_{1.97}$   \\
    Ni$_{0.95}$Re$_{0.05}$Te$_{2}$ & Ni$_{0.95}$Re$_{0.04}$Te$_{2}$ \\
   Ni$_{0.9}$Re$_{0.1}$Te$_{2}$ & Ni$_{0.92}$Re$_{0.099}$Te$_{2}$ \\
    Ni$_{0.8}$Re$_{0.2}$Te$_{2}$  & Ni$_{0.87}$Re$_{0.13}$Te$_{2}$  \\
    \hline
  \end{tabular}
\end{table}

Polycrystal samples of Ni$_{1-x}$Re$_{x}$Te$_{2}$ (x = 0, 0.05, 0.1, and 0.2) system were prepared by standard solid-state reaction method and single crystals of the same were grown by the modified Bridgman method. The Rietveld refinement was carried out considering the space group P-3m1 (164) having Ni and Te occupying 1a and 4d Wyckoff positions. CdI$_{2}$ prototype crystal structure of NiTe$_{2}$ has been shown in \figref{Fig1:crystal_structure}. Doping Re in Ni-site does not create extra peaks in the original NiTe$_{2}$ diffractogram for Ni$_{1-x}$Re$_{x}$Te$_{2}$ (x = 0, 0.05, and 0.1) whereas impurity peak is observed for x = 0.2 sample. The Rietveld refinement and RT XRD pattern on the naturally cleaved surface of the single crystal is shown in \figref{Fig2:xrd}. Single crystal x-ray diffraction and Laue image confirm the growth of crystal in [00n] direction.  The Energy-dispersive X-ray spectroscopy (EDS) analysis of the Ni$_{1-x}$Re$_{x}$Te$_{2}$ system is summarized in \tableref{tbl:EDS}. From RT XRD of polycrystal samples and EDS of single crystals it can be concluded that  Re was successfully doped in Ni- site for Ni$_{1-x}$Re$_{x}$Te$_{2}$ (x = 0, 0.05, and 0.1) whereas 13\% Re was successfully doped for x = 0.2.

\begin{figure}[htbp!]
\includegraphics[width=1.0\columnwidth]{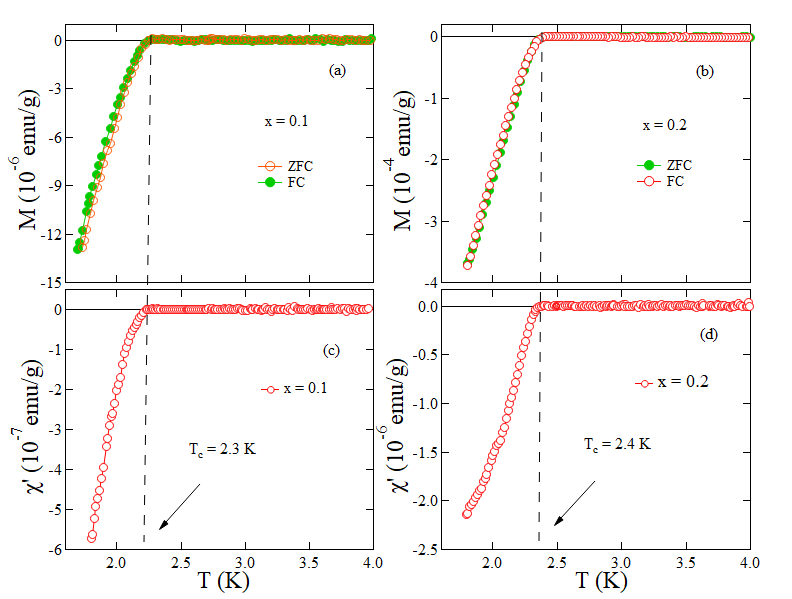}
\caption {\label{Fig3:superconducting transition} Temperature dependence of the dc (a, b) and ac (c, d) magnetic moment at 10 mT applied field for Ni$_{1-x}$Re$_{x}$Te$_{2}$ (x =  0.1, and 0.2).}
\end{figure}

\begin{figure}[htbp!]
\includegraphics[width=1.0\columnwidth]{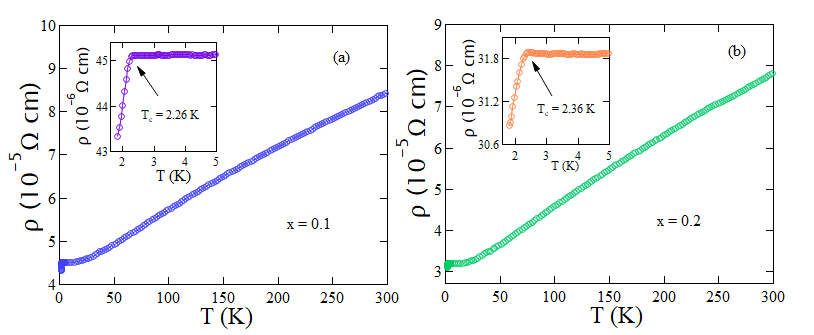}
\caption {\label{Fig4:resistivity} Temperature dependence of resistivity for Ni$_{1-x}$Re$_{x}$Te$_{2}$ (x =  0.1, and 0.2) whereas inset shows the sudden drop of resistivity.}
\end{figure}

\begin{figure}[htbp!]
\includegraphics[width=1.0\columnwidth]{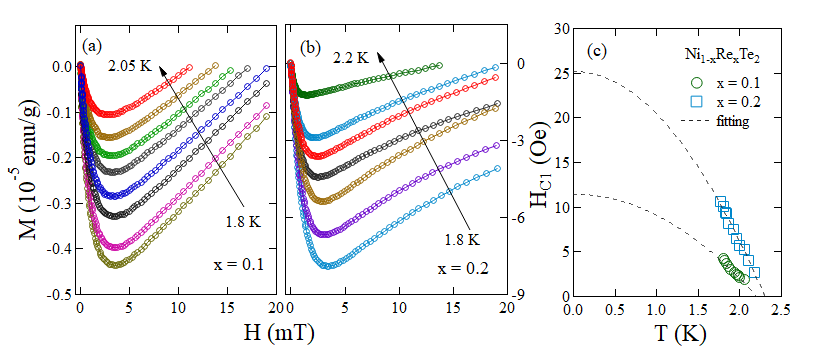}
\caption {\label{Fig5:magnetisation} Variation of magnetisation with field at different temperature for (a) Ni$_{0.9}$Re$_{0.1}$Te$_{2}$ and (b) Ni$_{0.8}$Re$_{0.2}$Te$_{2}$. (c) Lower critical field is fitted with GL equation.}
\end{figure}

\begin{figure}[htbp!]
\includegraphics[width=1.0\columnwidth]{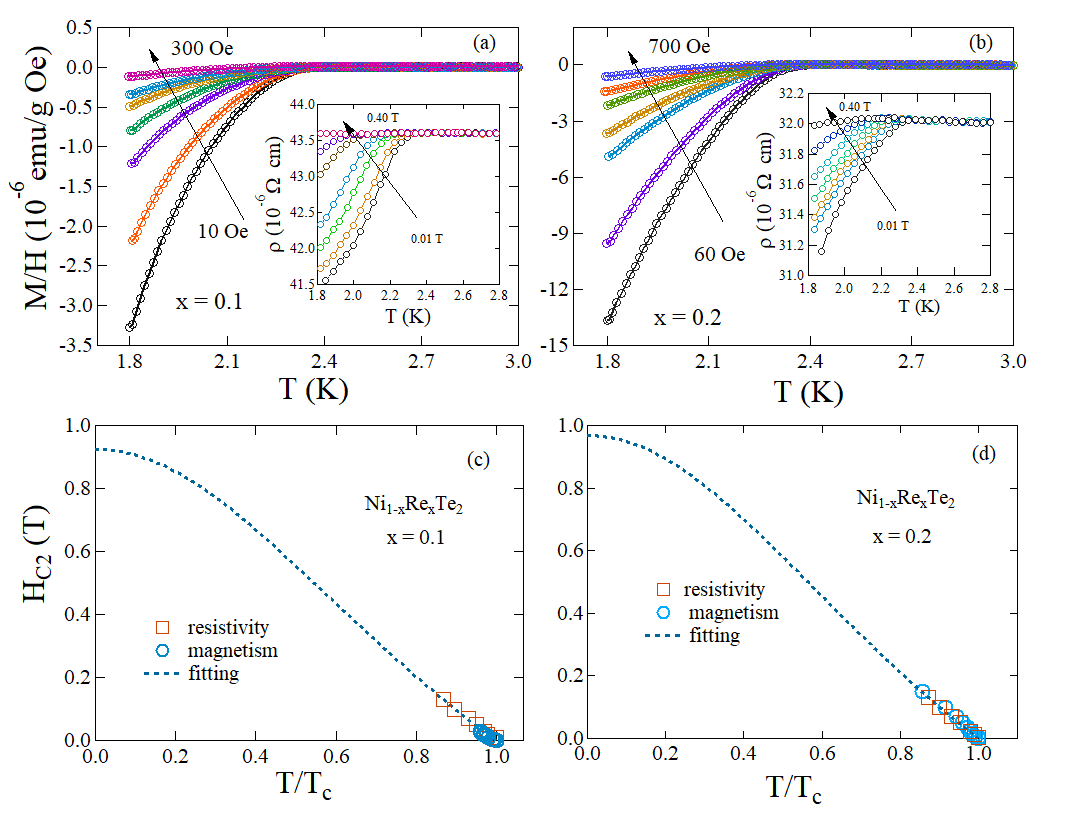}
\caption {\label{Fig6:Hc2} (a) and (b) Show the variation of magnetisation with temperature at different field and inset of the same figures show the variation of resistivity with temperature at different field for Ni$_{1-x}$Re$_{x}$Te$_{2}$ (x = 0.1, and 0.2). Higher critical field is fitted with GL equation shown in (c) and (d).}
\end{figure}

Magnetisation measurements were done for Ni$_{1-x}$Re$_{x}$Te$_{2}$ (x = 0, 0.05, 0.1 and 0.2) crystals. The parent NiTe$_{2}$ and Ni$_{0.95}$Re$_{0.05}$Te$_{2}$ sample do not show superconducting state down to 1.8 K, whereas the x = 0.1 and x=0.2 samples exhibit a clear signature of superconducting state at transition temperature, T$_{c}$ 2.3 K and 2.4 K respectively \figref{Fig3:superconducting transition}. Ac magnetisation measurements also show the same type of nature at 10 Oe field. The induced superconducting states are further verified by resistivity measurements. For both the samples x = 0.1 and 0.2 the sudden drop of resistivity confirm the evidence of superconductivity. The variation and the values of the T$_{c}$ are consistent with the magnetisation measurements (\figref{Fig4:resistivity}).
Magnetization measurements were done at different temperature for superconducting samples x = 0.1, and x = 0.2. The lower critical field, H$_{c1}$(0) was estimated from the magnetization curves in the range 0 to 200 Oe for both the samples in accordance with Ginzburg-Landau (GL) equation $H_{c1}(T)$ = $H_{c1}(0)\left(1-\left(\frac{T}{T_{c}}\right)^{2}\right)$. Experimental data were fitted  and calculated H$_{c1}$(0) is 11.45 and 25.24 Oe for x = 0.1 and 0.2 respectively, indicating a type-II superconductivity, shown in \figref{Fig5:magnetisation}. To rule out the possibility of superconductivity due to pure Re, which superconducting transition temperature can be varied (from 1.6K to 3.5K) by the preparation method of the specimen \cite{Re_1} as well as shear strain \cite{Re_2}. Magnetisation measurements of pure Re powder, shows type-I superconductivity at T$_{c}$ = 2.5 K with critical field 300 Oe. We have also annealed the Re followed by the same crystal growth procedure discussed in the experimental section to explore about annealing effect of Re powder on superconducting transition temperature. Magnetization measurements on annealed Re, which confirms type -I superconductivity at T$_{c}$ = 2 K with the low critical field (see supplementary information). The aforementioned results confirm that superconductivity in Ni$_{1-x}$Re$_{x}$Te$_{2}$ system is an intrinsic nature.
To study of the superconducting properties of the Ni$_{1-x}$Re$_{x}$Te$_{2}$ (x = 0.1, and 0.2) crystals in details, we have measured the resistivity and dc susceptibility under various applied magnetic fields (H). The gradual decrease of T$_{c}$ with the applied field for both the samples is shown in \figref{Fig6:Hc2}. The temperature variation of the upper critical field, H$_{c2}$(0) can be estimated by the Ginzburg Landau expression 
$H_{c2}(T)$ = $H_{c2}$(0)$\frac{(1-t^{2})}{(1+t^{2})}$, 
where $t$ = $\frac{T}{T_{c}}$. The calculated values of H$_{c2}$(0) are 0.92 T and 0.97 T for Ni$_{1-x}$Re$_{x}$Te$_{2}$ (x = 0.1, and 0.2 respectively). The superconducting characteristics parameters, Ginzburg Landau coherence lengths $\xi_{GL}$(0) and penetration depth, $\lambda_{GL}$(0) can be estimated from the relation 
$H_{c2}(0)$ = $\frac{\Phi_{0}}{2\pi\xi_{GL}^{2}}$ 
and $H_{c1}(0)$ = $\left(\frac{\Phi_{0}}{4\pi\lambda_{GL}^2(0)}\right)ln\left(\frac{\lambda_{GL}(0)}{\xi_{GL}(0)}+0.12\right)$, 
where $\Phi_{0}$ (= 2.07 $\times$10$^{-15}$ T m$^{2}$) is the magnetic flux quantum. The calculated values of  $\xi_{GL}$(0) are $\approx$ 189 \text{\AA} for Ni$_{0.9}$Re$_{0.1}$Te$_{2}$ and 184 \text{\AA} for Ni$_{0.8}$Re$_{0.2}$Te$_{2}$.

\begin{figure}[htbp!]
\includegraphics[width=1.0\columnwidth]{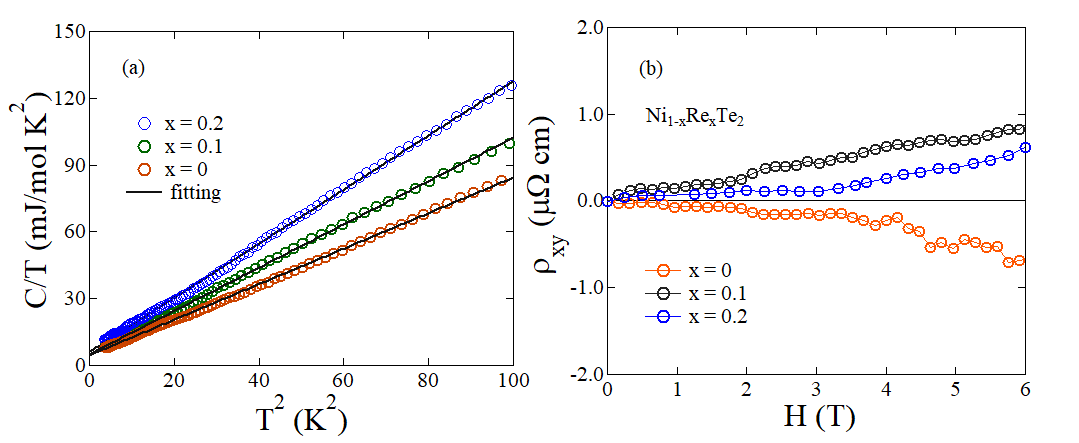}
\caption {\label{Fig7:sp} (a) The variation of C/T with T$^{2}$ in zero fields where the solid line is the fit in normal state. (b) Hall resistivity for Ni$_{1-x}$Re$_{x}$Te$_{2}$ (x = 0, 0.1 and 0.2).}
\end{figure}
We have performed the heat capacity and Hall measurements to analyze the electronic density of states and carrier concentration of the non-SC and SC samples \figref{Fig7:sp}. The low-temperature normal state-specific heat data were fitted using the equation $\frac{C}{T}$ = $\gamma+\beta_{3}T^{2}$ ($\gamma$= Sommerfeld coefficient and $\beta$ is the lattice contribution to the specific heat), which yields $\gamma$ = 4.59, 5.02 and 6.08 mJ/mol K$^2$  and $\beta_{3}$ = 0.80, 0.97, and 1.22 mJ/mol K$^4$ for x = 0, 0.1 and 0.2 respectively. On the basis of simple Debye model for the phonon contribution to the specific heat, we can calculate the Debye temperature by the expression, $\theta_{D}$ = $\left(\frac{12\pi^{4}RN}{5\beta_{3}}\right)^{\frac{1}{3}}$ where  N (= 3) is the number of atoms per formula unit, R is the molar gas constant (= 8.314 J mol$^{-1}$ K$^{-1}$). The estimated values of $\theta_{D}$ are 194, 181.63, and 168.45 K for x = 0, 0.1, and 0.2, respectively, which indicate the phonon softening in the lattice with Re doping in Ni-site. Considering the free electron model, the density of states values at Fermi level D$_{C}$(E$_{f}$) are calculated as 1.95. 2.13 and 2.56 states eV$^{-1}$ f.u.$^{-1}$ for x = 0, 0.1 and 0.2 respectively. The electron-phonon coupling strength ($\lambda_{e-ph}$) can be calculated using the McMillan formula. By considering the Coulomb pseudopotential $\mu^{*}$ = 0.1, $\lambda_{e-ph}$ are calculated as 0.54 and 0.56 for x = 0.1 and x = 0.2, respectively. These values indicate that the compounds are weak coupling superconductors. The substituting the Re in Ni site for NiTe$_{2}$ facilitates the enhancement of the electron-phonon coupling and DOS at the Fermi energy. The Hall measurement data are taken at temperature 10 K for x = 0, 0.1, and 0.2 samples. The slope is negative for undoped sample, but it changes to positive for x = 0.1 and 0.2 samples, which suggests a change in the sign of carrier types \cite{NbMoTe2}. This is the evidence for hole-like charge carriers in Ni$_{1-x}$Re$_{x}$Te$_{2}$ (x = 0.1 and 0.2) sample with a strong contrast to that of parent NiTe$_{2}$ where the conducting carriers are found to be electron-type. This, in fact, indicates that a significant reconstruction of the Fermi surface is occurring with Re substitution which facilitates the superconductivity similar to hole-doped superconductivity in (La$_{1-x}$Sr$_{x}$)OFeAs \cite{La}, (Sr$_{1-x}$K$_{x}$)Fe$_{2}$As$_{2}$ \cite{Sr}. However, further experiments are required to gain deeper insights into Fermi surface reconstruction.  A summary of all the experimentally measured and estimated parameters is given in \tableref{tbl:parameters}.
\begin{table}
  \caption{Normal and superconducting parameters of Ni$_{1-x}$ Re$_{x}$Te$_{2}$ samples.}
  \label{tbl:parameters}
  \begin{tabular}{lllll}
    \hline
Parameters & unit & x = 0 & x = 0.1 & x = 0.2 \\
\hline
\\[0.5ex]                              
T$_{c}$& K & & 2.3 &  2.4\\
H$_{c1}(0)$& Oe & & 11.45 & 25.24\\ 
H$_{c2}$(0)& T & & 0.92 & 0.97\\
$\xi_{GL}$& \text{\AA}& & 189 &  184\\
$\gamma_{n}$& mJ mol$^{-1}$ K$^{-2}$& 4.59 & 5.02 &  6.08\\
$\beta_{3}$& mJ mol$^{-1}$ K$^{-4}$& 0.80 & 0.97 &  1.22\\
$\theta_{D}$& K& 194 & 182 & 168\\
$\lambda_{e-ph}$&  & & 0.54 & 0.56\\
D$_{C}$(E$_{f}$) & states/eV f.u.& 1.95 & 2.13 & 2.58\\
\\[0.5ex]
\hline\hline
\end{tabular}
\end{table}

\section{Conclusion}
In summary, we have successfully prepared a series of layered Ni$_{1-x}$Re$_{x}$Te$_{2}$ crystals with nominal x = 0 to 0.20 by modified Bridgman method having space group  P-3m1 (164).  The magnetic and transport measurements highlight that Re substitution in Ni-site in NiTe$_{2}$ compound induces superconductivity, although the bulk crystal is not known as a superconductor. Our main results demonstrate unambiguously that Ni$_{1-x}$Re$_{x}$Te$_{2}$ (x = 0.1 and 0.2) displays type-II superconductivity at T$_{c}$ = 2.3 K  and 2.4 K with higher critical field H$_{c2}$(0) 0.92 and 0.97 T respectively. The specific heat measurements affirm that electron-phonon coupling constant $\lambda_{e-ph}$ and the density of states D$_{C}$(E$_{f}$) increases with the increment of Re content. Evidence for hole-like charge carriers has been illustrated by Hall effect measurements in Re doped sample in the contrast to undoped compound where an electron is the main charge carrier. The estimated values of Debye temperature, $\theta_{D}$ indicate the phonon softening in the lattice with Re doping in Ni-site, which is consistent with hole doped superconductors. Therefore, our finding of superconductivity in the hole-doped side will widely open the territory for exploring new 2D superconductors with the topological states in 2D van der Waals materials.

\section{Acknowledgement}
R.~P.~S.\ acknowledges Science and Engineering Research Board, Government of India for the Core Research Grant CRG/2019/001028.

\bibliography{NTR}
\bibliographystyle{h-physrev}

\begin{figure}[htbp!]
\includegraphics[width=1.0\columnwidth]{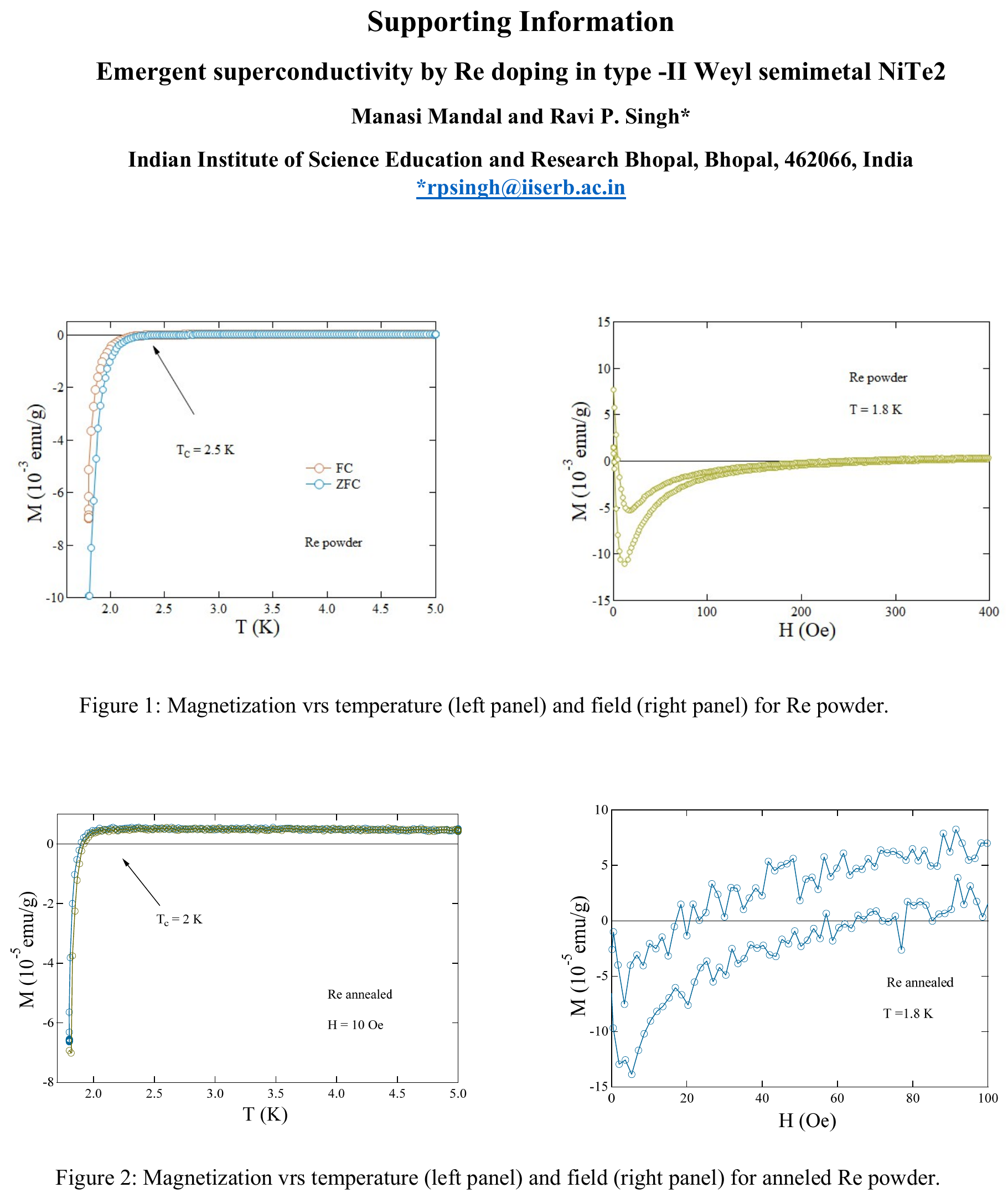}
\caption {\label{Fig1:crystal_structure} CdI$_{2}$ prototype crystal structure of NiTe$_{2}$ having the space group P-3m1 (164).}
\end{figure}

\end{document}